\documentclass[sigconf]{acmart}

\usepackage{hyperref}
\usepackage{cleveref}
\usepackage{xurl}

\usepackage{booktabs}
\usepackage{tabularx}
\usepackage{enumitem}

\usepackage{fontawesome5}

\usepackage{mathtools}
\usepackage{MnSymbol} 

\AtBeginDocument{%
  }

\setcopyright{acmlicensed}
\copyrightyear{2026}
\acmYear{2026}
\acmDOI{XXXXXXX.XXXXXXX}

\acmConference[EASE 2026]
    {The 30th International Conference on Evaluation and Assessment in Software Engineering}
    {Tue 9 - Fri 12 June 2026}
    {Glasgow, United Kingdom}
    
\acmISBN{978-1-4503-XXXX-X/2018/06}

\begin{document}

\title{Towards Improving the External Validity of Software Engineering Experiments with Transportability Methods}

\author{Julian Frattini}
\orcid{0000-0003-3995-6125}
\author{Richard Torkar}
\orcid{0000-0002-0118-8143}
\additionalaffiliation{
  \institution{The Stellenbosch Institute for Advanced Study}
  \city{Stellenbosch}
  \country{South Africa}
}
\author{Robert Feldt}
\orcid{0000-0002-5179-4205}
\additionalaffiliation{
  \institution{Mid Sweden University}
  \city{Östersund}
  \country{Sweden}
}
\email{{firstname}.{lastname}@chalmers.se}
\affiliation{
  \institution{Chalmers University of Technology and University of Gothenburg}
  \city{Gothenburg}
  \country{Sweden}
}

\author{Carlo A. Furia}
\orcid{0000-0003-1040-3201}
\email{furiac@usi.ch}
\affiliation{
  \institution{USI Università della Svizzera italiana}
  \city{Lugano}
  \country{Switzerland}}

\renewcommand{\shortauthors}{Frattini et al.}

\begin{abstract}
    Controlled experiments are a core research method in software engineering (SE) for validating causal claims.
    However, recruiting a sample of participants that represents the intended target population is often difficult or expensive, which limits the external validity of experimental results.
    At the same time, SE researchers often have access to much larger amounts of observational than experimental data (e.g., from repositories, issue trackers, logs, surveys and industrial processes).
    \emph{Transportability methods} combine these data from experimental and observational studies to ``transport'' results from the experimental sample to a broader, more representative sample of the target population.
    Although the ability to combine observational and experimental data in a principled way could substantially benefit empirical SE research, transportability methods have---to our knowledge---not been adopted in SE\@. 
    In this vision, we aim to help make that adoption possible.
    To that end, we introduce transportability methods and their prerequisites, and demonstrate their potential through a simulation.
    We then outline several SE research scenarios in which these methods could apply, e.g., how to effectively use students as substitutes for developers. 
    Finally, we outline a road map and practical guidelines to support SE researchers in applying them.
    Adopting transportability methods in SE research can strengthen the external validity of controlled experiments and help the field produce results that are both more reliable and more useful in practice.
\end{abstract}



\received{23 January 2026}
\received[accepted]{2 April 2026}

\maketitle

\section{Introduction}
\label{sec:intro}

Controlled experiments are an essential research method in software engineering (SE)---as in any empirical discipline---for validating claims about causal relationships between variables~\cite{wohlin2012experimentation}.
The random assignment of study subjects to a treatment or control group eliminates the influence of confounding factors on the relationship of interest~\cite{shadish2002experimental}.
Therefore, the observed effect can be attributed to the treatment rather than to confounding effects.

However, this internal validity often comes at the expense of external validity.
Experiments are conducted in a contrived setting~\cite{stol2018abc} where a representative sample of subjects must be drawn from a target population~\cite{shadish2002experimental,sjoberg2005survey}.
Achieving a broad and representative sample is particularly challenging when an experiment involves human subjects: 
the intended target population of SE professionals is difficult to reach and expensive to recruit~\cite{baltes2022sampling}.
Consequently, controlled experiments in SE often settle for small samples and participants with a limited experience (e.g., students), thus jeopardizing statistical power and external validity~\cite{dybaa2006systematic}.
This hinders transfer of scientific results into practice~\cite{sjoberg2003challenges}.

Other empirical disciplines face the same challenges. 
For example, medical researchers aim to predict how well a treatment response observed in a sample will hold in the target population, i.e., all potential recipients of that treatment~\cite{rothwell2005external}.
To address this, the field of statistical causal inference has developed a formal framework for \emph{transportability} of statistical relations across populations~\cite{pearl2011transportability}.
Within this broader line of work, ``[e]stimation methods to generalize trial findings to a target population of interest''~\cite{colnet2024causal} emerged, which we will refer to as \textit{transportability methods} from here on out. 
These methods combine experimental results with typically much larger observational data on relevant covariates, allowing to transport results from limited experiments to a target population without collecting more experimental data~\cite{colnet2024causal}. 

Despite their potential, such methods have---to our knowledge---not been adopted in SE research to date.
With this vision paper, our goal is to pave the way for the adoption of transportability methods in SE research by contributing: (1) a high-level description of transportability methods and their necessary preconditions (\Cref{sec:transport}), (2) a simulation demonstrating the methods' usefulness (\Cref{sec:simulation}), (3) a list of valuable cases for application in SE research (\Cref{sec:cases}), and (4) a road map for enabling adoption in SE research (\Cref{sec:roadmap}).


\subsection*{Data Availability Statement}
All figures, scripts, and documentation can be found in our replication package~\cite{replication-package}.

\section{Transportability Methods}
\label{sec:transport}

\Cref{sec:transport:scenario} introduces an illustrative example contextualizing the subsequent, methodological descriptions.
Then, \Cref{sec:transport:preconditions} lists relevant preconditions that need to be met for the actual transportability methods described in \Cref{sec:transport:methods} to work.

\subsection{Illustrative Example}
\label{sec:transport:scenario}

At a high level, a controlled experiment estimates the causal effect of a treatment $A$ on an outcome $Y$ (i.e., $A \rightarrow Y$). 
As a running example, we will consider the effect of using (i.e., the treatment $A = 1$) or not using (i.e., the control $A = 0$) generative AI (GenAI) on the number $Y$ of successfully identified defects during code reviews~\cite{TufanoMTDHB25}.
The quantity of interest to estimate from the experiment is the \textit{average treatment effect} (ATE) $\tau$, i.e., the average difference in detected defects when using GenAI instead of not using it.

\begin{figure}
    \centering
    \includegraphics[width=0.9\linewidth]{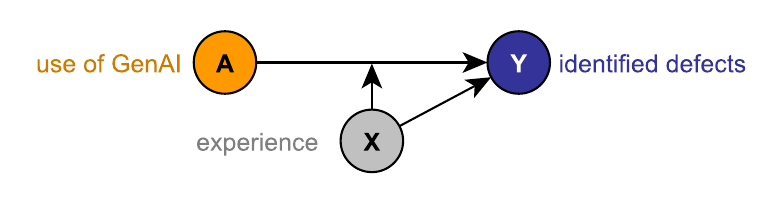}
    \vspace{-0.5cm}
    \caption{DAG visualizing causal assumptions of the illustrative, running example}
    \label{fig:dag}
\end{figure}

The level of experience $X$ of a subject is an example of a covariate that may affect the outcome $Y$ directly ($X \rightarrow Y$), but may also \emph{moderate} the ATE~\cite{siegmund2015confounding}:
Subjects with less experience may benefit more from using GenAI during code reviews than subjects with more experience. 
In the absence of experience, suggestions from GenAI may be a decent help, while the same suggestions may be trivial for an experienced reviewer.
This makes $X$ a \emph{treatment effect modifier}.
\Cref{fig:dag} visualizes these relations as a directed acyclic graph (DAG), commonly used in Pearl's framework for causal inference~\cite{pearl2011transportability}.

\begin{figure}
    \centering
    \includegraphics[width=\linewidth]{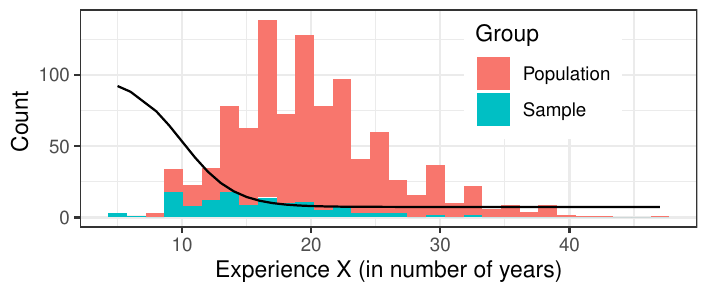}
    \vspace{-0.7cm}
    \caption{Distributions of the covariate $X$ in the target population and experimental sample due to covariate shift}
    \label{fig:sampling}
\end{figure}

Controlled experiments aim to approximate the ATE $\tau$ in the target population, but can realistically only measure the \emph{trial} ATE $\tau_1$ in the experimental sample.
The ATE of interest $\tau$ may differ from the measurable trial ATE $\tau_1$.
In our illustrative example, one reason for this difference may stem from the challenge of recruiting subjects.
In particular, \textit{trial eligibility} $S$ (i.e., the likelihood of a subject from the target population to be included in the experimental sample) is often affected by a treatment effect modifier such as $X$:
We can assume that subjects with more experience $X$ are in more senior positions because it increases the likelihood of getting promoted~\cite{waldman1984worker}.
This makes them less accessible and more expensive to recruit as subjects to the experiment.
In contrast, subjects with less experience $X$ may be more available to participate in the experiment.
For this reason, SE experiments commonly use university students to represent the target population of software engineers~\cite{carver2004issues}.
\Cref{fig:sampling} visualizes this challenge.
The black line represents subjects' trial eligibility (in percentage) which decreases for higher values of $X$.
The distribution of the covariate $X$ in the experimental sample (teal bars) therefore ends up different from the distribution in the target population (red bars)---a phenomenon known as \textit{covariate shift}~\cite{sugiyama2012machine}.
When $X$ is both a treatment effect modifier and experiences a covariate shift, the measurable $\tau_1$ can differ from $\tau$.
In our example, the experiment likely involves more easy-to-recruit but inexperienced subjects for which the measured effect is particularly strong.
As a consequence, the experiment will overestimate $\tau_1 > \tau$ and suggest that using GenAI is much more effective than it would be in the target population.

\subsection{Preconditions}
\label{sec:transport:preconditions}

\Cref{tab:assumptions} lists the preconditions that must hold for the transportability methods (presented in \Cref{sec:transport:methods}) to work, as elicited by Colnet et al.~\cite{colnet2024causal}.
Preconditions A1 through A4 are fundamental requirements for a valid controlled experiment.
Preconditions A5, A6, and A7 are specific to transportability methods.
Thus, we discuss what they mean and what happens if they are violated in the following.

\begin{table*}[hbt]
    \footnotesize
    \centering
    \caption{Assumptions to apply transportability methods to controlled experiments. An asterisk * marks those that are specific to transportabilty techniques.}
    \label{tab:assumptions}
    \begin{tabularx}{\linewidth}{lp{2.5cm}|p{3.05cm}|X}
        \toprule
        \textbf{ID} & \textbf{Name} & \textbf{Definition} & \textbf{Explanation} \\
        \midrule
        A1 & Consistency & $Y = AY(1) + (1-A)Y(0)$ & The observed outcome is the potential outcome given the assigned treatment, i.e., we have a connection between treatment and outcome. \\
        A2 & Randomization & $\left\{ Y(0), Y(1) \right\} \upmodels A \mid S = 1, X$ & The treatment is independent of all the potential outcomes and covariates as in a controlled experiment. \\
        A3 & Ignorability on trial participation & $\left\{ Y(0), Y(1) \right\} \upmodels S \mid X$ & The outcome $Y$ is unaffected by trial participation $S$ when controlling all relevant covariates $X$. \\
        A4 & Mean exchangeability & $\mathbb{E}\left[ Y(a) \mid X=x, S=1 \right] =$ $\mathbb{E}\left[ Y(a) \mid X=x \right]$ & Instead of requiring that every individual behaves identically in and out of the trial, we only assume that \textit{on average} the treatment effect is the same between groups with the same observed characteristics. \\
        A5* & Sample ignorability for treatment effects & $Y(1) - Y(0) \upmodels S \mid X$ & The effect of the treatment is independent of trial eligibility when knowing all covariates $X$. \\
        A6* & Transportability of the conditional ATE & $\forall x \in X \colon \tau_1(x) = \tau(x)$ & In every stratum $x \in X$, the ATE in the experimental sample $\tau_1$ is equal to the ATE in the target population $\tau$. \\
        A7* & Positivity of trial participation & $\exists c \colon \mathbb{P}(S = 1 \mid X ) \geq c$ &  Every subject in the target population must have at least some chance (i.e., non-zero probability) of being included in the experiment. \\
        \bottomrule
    \end{tabularx}
\end{table*}

A5 requires that at least one covariate $X$ affects trial eligibility $S$, i.e., the black line in \Cref{fig:sampling} is not just a horizontal line.
If A5 does not hold, then the distribution of $X$ would be the same in the experimental sample as in the target population.
In such a case, the experimental sample perfectly represents the target population (i.e., there is no covariate shift), the trial ATE would perfectly generalize ($\tau = \tau_1$), and there would be no need for transporting.
As we argued in the illustrative example in \Cref{sec:transport:scenario}, and as we will further elaborate in \Cref{sec:cases}, in most practical cases there would be at least some covariate shift, i.e., A5 normally holds.

A6 requires that for every stratum of $x \in X$ the ATE in the target population is the same as the trial ATE, i.e., the \textit{conditional} ATE is the same even if the marginal ATE may not be.
In other words, if we stratify by the covariate $X$, the trial ATE generalizes to the target population.
This implies that if A6 holds, $X$ acts as the only treatment effect modifier.
A situation where A6 would not hold is if there are other \emph{unobserved} covariates that moderates the treatment effect.
In this case, applying transportabilty methods to only $X$ may fail to correct for all of the covariate shift.

Finally, A7 requires that every subject from the target population has non-zero probability of being included in the experimental sample, i.e., the black line in \Cref{fig:sampling} is always above 0\%.
In general, the distribution of $X$ in the experimental sample will differ from in the target population (see A5).
A7 only requires that the two distributions have the same support.
If A7 does not hold---i.e., some stratum of $X$ has a 0\% probability of being sampled---no statistical method could recover the ATE from the unobserved stratum.
Transportability is, hence, constrained to the range of $X$ covered in the experimental sample.

\subsection{Formulae and Estimation Methods}
\label{sec:transport:methods}

One approach to approximate the ATE $\tau$ is to model the treatment effect modification as an interaction effect in a regression formula:
\begin{equation}
    Y \sim \mathcal{N}(\alpha + \tau \cdot A \cdot X, \epsilon)
    \label{eq:linreg}
\end{equation}
This formula regresses the outcome $Y$ (here assumed to be normally distributed $\mathcal{N}$ with variance $\epsilon$) on a linear combination of an intercept $\alpha$ (the baseline value for $Y$) and the treatment $A$, which has an effect of $\tau$ on the outcome but is moderated by $X$.
For simplicity of the demonstration, we ignore all marginal effects of $A$ and $X$ on $Y$.

However, this approach of estimating $\tau$ only works if the interaction between the continuous $X$ and $A$ is linear.
This would require that every increase in the covariate $X$ causes the same proportional increase in the treatment effect moderation.
However, not every effect behaves this way, particularly when considering human factors~\cite{li2018curvilinear}.
In the example where the covariate $X$ is a continuous measure of experience in number of years, it is possible that an increase of experience from 0 to 1 year has a greater effect than from 20 to 21 years.
To handle these more complex interactions, a more general approach is needed.

Enter transportability methods.
Under the conditions in \Cref{sec:transport:preconditions}, a transportability method can recover the actual ATE $\tau$ of the target population from (1) the trial ATE $\tau_1$ and (2) the distribution the the covariates $X$ in the target population, but without requiring further data about $A$ or $Y$.

Colnet et al. discuss two classes of identification formulae~\cite{colnet2024causal}:

\begin{enumerate}[leftmargin=0.5cm]
    \item \textbf{Reweighting}: $\tau=\mathbb{E} \left[ \frac{n}{m \times \alpha(X) } \tau_1(X) \mid S=1 \right]$
    \item \textbf{Regression}: $\tau = \mathbb{E}\left[\mu_{A=1,S=1}(X)-\mu_{A=0,S=1}(X)\right] = \mathbb{E}\left[\tau_1(X)\right]$
\end{enumerate}

Based on these formulae, they elaborate several estimation methods for transportability.
For brevity, we will only present one from each class and refer the interested reader to Colnet et al.~\cite{colnet2024causal}.

\paragraph{Transport with reweighting:}
The \textbf{inverse probability of sampling weighting} (IPSW) is an estimator of $\tau$ based on reweighting. 
IPSW weighs each data point in the controlled experiment based on trial eligibility:

\begin{equation}
    \hat{\tau}_{\text{IPSW}} = \frac{1}{n}\sum^n_{i=1} \frac{n}{m} \frac{Y_i}{\hat{\alpha}_{n,m}(X_i)} \left(\frac{A_i}{e_1(X_i)} - \frac{1-A_i}{1-e_1(X_i)} \right)
    \label{eq:ipsw}
\end{equation}

Here, $n$ is the size of the experimental sample, $m$ the size of the larger target population, $\hat{\alpha}_{n,m}(X_i)$ represents the trial eligibility $S$, and $e_1(x)$ the propensity score~\cite{colnet2024causal} (i.e., the likelihood of being assigned to a treatment, which is fixed at 50\% in most experiments with only one treatment and one control level).
Trial eligibility $\hat{\alpha}_{n,m}(X_i)$ can be estimated via logistic regression based on the distribution of $X$ in the experimental sample and in the target population.
Values of $X$ that occur often in the sample and in the target population have a high trial eligibility, values of $X$ that occur rarely in the sample but more often in the target population have a low trial eligibility.
Based on this estimated trial eligibility, the data points from the controlled experiment are re-weighted.
Data points from subjects with high trial eligibility contribute less to the ATE than from subjects with low trial eligibility.
In the illustrative example, this would mean that the results obtained from one participating senior engineer (high experience $X$, and therefore, low trial eligibility) are weighted more strongly in estimating the ATE than the results obtained from several participating master students (low experience $X$, and therefore, high trial eligibility).
This weighting of results by the inverse probability of sampling counteracts the effect of the covariate on the trial eligibility.

\paragraph{Transport with regression:}
The \textbf{plug-in g-formula} is an estimator of $\tau$ based on regression. 
This estimator approximates $\tau$ by fitting two separate linear models.

\begin{equation}
    \hat{\tau}_{G} = \frac{1}{m}\sum_{i = n+1}^{n+m} \left(\hat{\mu}_{1,1,n}(X_i) - \hat{\mu}_{0,1,n}(X_i)\right)
    \label{eq:gform}
\end{equation}

The two linear models predict the outcome $Y$ based on $X$, one for the control group ($\hat{\mu}_{0, 1}$) and one for the treatment group ($\hat{\mu}_{1, 1}$).
These regressions $Y \sim X$ for the two levels of $A$ directly model the treatment effect moderation of $X$.
The plug-in g-formula estimation then applies the covariate value $X_i$ of all $m$ observational data points to both linear models, averages the results, and calculates the ATE as the difference between the two averages.

\section{Simulation}
\label{sec:simulation}

To demonstrate how transportability methods work in practice, we perform a computer simulation~\cite{stol2018abc}.
We simulate a target population and draw a sample from it that represents participants of an experimental study.
We then simulate this experiment with known causal effects among variables.
Finally, we estimate the ATE using four methods: mean difference, linear regression with an interaction term, and the two presented transportability methods.
We compare the four methods in their ability to recover the simulated causal effect from the data.

\subsection{Dataset construction}

We simulate the illustrative example described in~\Cref{sec:transport:scenario}.
The main factor $A$ has two levels: control ($A=0$, i.e., not using AI) and treatment ($A=1$, i.e., using AI).
We use a normally distributed measure representing \textit{defect detection performance} instead of the number of identified defects for the outcome $Y \in \mathbb{R}$.
Using this normally distributed outcome simplifies interpretation by avoiding link functions required for count data~\cite{mcelreath2018statistical}, though the transportability methods work just as well for such data.
Finally, the covariate $X \in \mathbb{R}^+$ represents experience measured in number of years and follows a negative-binomial (NB) distribution (as in \Cref{fig:sampling}). 


We created a data set by first simulating a target population of 1000 subjects with a random distribution of the covariate $X \sim \text{NB}(10, 3)$.
The scale parameter $\mu = 10$ and dispersion parameter $\gamma = 3$ are arbitrary but produce realistic values between 0 and about 50 years of experience with a peak around $X=20$, as seen in \Cref{fig:sampling}.
Next, we simulated the trial eligibility $S \sim \mathrm{Bernoulli}(p)$, where the likelihood of being included in the experimental sample decreases with $X$ (as shown as the black line in \Cref{fig:sampling}).
Subjects where $S=1$ are included in the controlled experiment, the remaining subjects where $S=0$ remain in the observational group.
This split the data set into roughly $n = 175$ experimental subjects and $m = 825$ observational subjects, though the exact numbers vary due to the random distribution.
Finally, we randomly divided the experimental subjects into control ($A=0$) and treatment ($A=1$) groups and simulated the outcome $Y$ which is affected by the treatment $A$ but moderated by the covariate $X$.
For the ATE, we chose an arbitrary value of $\tau = 16.7$.
For the treatment effect moderation, our simulation decreased the ATE with higher values of $X$ in a non-linear way, which models diminishing returns of increasing experience.
We did not simulate a marginal effect of $X \rightarrow Y$, i.e., the outcome $Y$ did not change for different values of $X$ directly, only through the treatment effect moderation.

\subsection{Estimation Setup}

In the evaluation, we compare four methods to estimate the ATE:

\begin{enumerate}
    \item \textbf{Mean difference} (baseline) between the outcome $Y$ in the control and treatment group
    \item \textbf{Linear regression model} with an interaction effect representing the treatment effect moderation (\Cref{eq:linreg})
    \item \textbf{IPSW estimator} from the reweighting-class (\Cref{eq:ipsw})
    \item \textbf{Plug-in g-formula} from the regression-class (\Cref{eq:gform})
\end{enumerate}

We run the simulation described above 50 times. 
For each simulated dataset, we record the ATE estimated by each of the four methods and then plot the distribution of these estimates.

\subsection{Results}

\begin{figure}
    \centering
    \includegraphics[width=\linewidth]{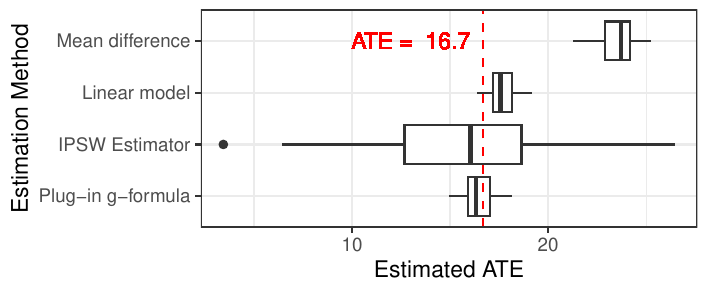}
    \vspace{-0.7cm}
    \caption{Results from the simulation compared against the simulated effect (red line).}
    \label{fig:sim:results}
\end{figure}

\Cref{fig:sim:results} shows the results of the simulation.
The box plots represent the estimated results of each of the four methods over 50 iterations.
The red dashed line shows the simulated ATE ($\tau = 16.7$) that these methods attempted to recover.

The na\"ive mean difference vastly overestimates the simulated ATE.
Since the experimental sample predominantly contained subjects with lower experience $X$ and the ATE of the main factor $A$ is moderated to be stronger for lower values of $X$, the na\"ive estimation assumes the ATE to be much stronger than it truly is ($\tau_1 \gg \tau$).

The linear model including an interaction effect performs significantly better, but still overestimates the simulated ATE.
This is because it models the interaction to be linear, while the treatment effect moderation is actually non-linear.

The 50\%-quantiles of estimations of both transportability methods include the simulated ATE thanks to the covariate distribution $X$ in the target population.
However, the IPSW estimator shows substantially greater uncertainty around its mean estimate.
As Colnet et al. explain, this estimator can be highly unstable, particularly when the trial-eligibility weights become extreme~\cite{colnet2024causal}.
The plug-in g-formula performs better on both accounts: it is more accurate and more robust.
Using the information about the covariate distribution from the target population, it is able to correct the treatment effect moderation and recover the simulated ATE.

\section{Motivating Examples}
\label{sec:cases}

Beyond the illustrative example used in \Cref{sec:transport,sec:simulation}, we identify three classes of challenges in empirical SE research for which transportability methods may be worth considering.

\subsection{Experiment Participant Experience}
\label{sec:cases:subjects}

A long-running debate in SE research asks whether (undergraduate) students can serve as valid substitutes for SE professionals in controlled experiments~\cite{curtis1986way,salman2015students}. 
Students are easier to recruit, but they may lack the skills or domain knowledge of professional practitioners~\cite{dieste2013software}.
This question has fueled an extensive public discussion, with prominent empirical SE researchers arguing both sides~\cite{carver2004issues,falessi2018empirical,feldt2018four}.
Yet the debate has relied mostly on hypotheses, assumptions, and anecdotal evidence rather than direct empirical tests.
Transportability methods offer a constructive way forward for understanding and addressing the issue of representative subjects in SE experiments.

\subsection{System Properties}
\label{sec:cases:objects}

Not only human participants but also the artifacts used in experiments may fail to represent the target population.
Researchers often study software systems built in student projects~\cite{hey2024requirements}, specifications mocked for the experiment~\cite{frattini2025applying}, or artificial bugs injected into software~\cite{JustJIEHF14}.
Industry-grade artifacts may be unavailable, unsuitable for time-constrained experiments, or missing properties that the study requires (e.g., ground-truth traceability links~\cite{hey2024requirements}).
Even when experiments use industry-grade artifacts, they are often restricted to open-source systems because those are accessible~\cite{sens2024large}.

Smaller, simpler, hand-crafted, or open-source artifacts are often more practical, but they may not represent the target population of software systems, specifications, or other artifacts.
This creates covariate shift in characteristics such as size, complexity, documentation quality, which affects how well results generalize.
Framed as a transportability problem, the objects' representativeness becomes a tangible property and limitations to external validity clear.

\subsection{Task Complexity}
\label{sec:cases:tasks}

In addition to human and artifact subjects, the experimental tasks themselves are often not fully representative of real-world practice~\cite{sjoberg2005survey}.
Researchers often limit the scope of a task to minimize the required time commitment of participants, e.g., code reviews without extensive familiarization with the source code~\cite{TufanoMTDHB25}.
This sacrifices representativeness of a task, raising the question whether effects observed during the experimental task still hold in reality.

\section{Road Map and Guidelines}
\label{sec:roadmap}

Even in medical research, where transportability methods originated, their application is still limited~\cite{colnet2024causal}.
We see the opportunity to enable this useful class of methods for SE research by focusing effort on the following steps.

\subsection{Understanding Treatment Effect Modifiers}
\label{sec:roadmap:understanding}

Firstly, SE research should develop a clear understanding of which covariates act as moderators on ATEs of interest, as these limit external validity of experimental results.
Identifying them would support a more rigorous and systematic analysis of the threats to external validity, rather than resorting to common practice~\cite{wyrich2024evidence}.

We anticipate that several covariates will be specific to certain SE tasks, while others apply to a broader scope.
For example, \textit{experience}, \textit{domain knowledge}, or \textit{skill} probably moderate many causal effects of interest~\cite{wagner2021code}, as they are likely to influence almost any SE activity.
In contrast, a covariate like \textit{programming language proficiency} will affect some SE tasks (e.g., source code development and code reviews)~\cite{TufanoMTDHB25} more than others (e.g., requirements elicitation).

Although identifying all moderators is difficult, causal models (\Cref{fig:dag}) make these assumptions explicit. 
Rather than aiming for ``perfect'' knowledge, researchers should use these models for sensitivity analyses that quantify how unobserved moderators could bias the transported ATE~\cite{mcelreath2018statistical}.
This shifts the focus from exhaustive completeness to the statistical robustness of the external-validity claim.
These analyses can also help researchers prioritize the factors that matter most when designing an experiment: 
They should collect data on key moderating covariates and seek a representative sample that spans the full range of each covariate.

\subsection{Operationalizing Covariates}
\label{sec:roadmap:operationalizing}

Once relevant covariates are identified, SE research must develop appropriate and agreed-upon operationalizations.
Since many of the moderating covariates are likely to be latent variables and context factors, their operationalization is critical~\cite{sjoberg2022construct}.
For example, \textit{experience} is often operationalized via the \textit{number of years working as a software engineer}, which may not adequately reflect the underlying concept:
If one software engineer has worked for twice as long as another, there is no guarantee that they are also ``twice as experienced.''
A proper operationalization underpins the construct validity of these covariates.
Without it, the previously introduced transportability methods lose effectiveness. 
Therefore, thoroughly assessing the construct validity of operationalizations of covariates moderating an ATE~\cite{terwee2018cosmin} will pave the way towards adjusting for them using transportability methods.

\subsection{Collecting Observational Data}
\label{sec:roadmap:collecting}

With relevant covariates identified and operationalized, the SE research community can steer its efforts towards collecting observational data on these covariates in the target population.
While surveying the total target population remains unrealistic, observational studies collecting covariate distributions are likely to involve larger samples of the target distribution compared to interventional studies (e.g., controlled experiments or action research studies), given that they are less obtrusive~\cite{stol2018abc}.
For example, if \textit{experience} is identified as a relevant, ATE-moderating covariate for several SE tasks, surveys collecting the distribution of developer experience in different countries and companies can be conducted to approximate the distribution of that covariate in the general target population.

\subsection{Transporting Results}
\label{sec:roadmap:transporting}

With observational data sets approximating the distribution of relevant covariates, SE researchers can transport the results of controlled experiments from an experimental to an observational sample, where the latter is more representative of the target population.
Thanks to the previously presented methods, covariate shift in controlled experiments can be partially addressed when the assumptions hold.
For example, controlled experiments can be conducted primarily with students (i.e., subjects with lower \textit{experience}) as long as there are still a few subjects representing the other end of the spectrum of the covariate (i.e., subjects with higher \textit{experience}) to meet A7.
This also implies the advice that---given an existing sample of students---effort is better spent on recruiting a few senior software engineers instead of a lot more students.
Ultimately, when an experiment meets the assumptions in \Cref{sec:transport:preconditions} and observational data on ATE-moderating covariates is available, transportability methods can improve the external validity of results without requiring additional experimental data.

\subsection{Presenting Results}
\label{sec:roadmap:presenting}

Finally, these methods allow contextualizing obtained results in two regards.
First, the presence of a treatment effect modifier allows complementing the ATE with the results about the actual moderation.
While the ATE represents the \emph{average} effect aggregated over the full range of the covariate $X$, a stratified view into how the effect changes along $X$ provides more detailed insights.
In the illustrative example, this would allow the conclusion that the use of GenAI is beneficial for inexperienced subjects but irrelevant for experienced ones.
Second, assessing the degree to which precondition A7---the positivity of trial participation---is met allows confining the external validity of the achieved results.
If it was impossible to recruit subjects or infeasible to sample objects that cover the full spectrum of a treatment effect modifying covariate, the obtained range should be reported to confine the scope of generalizability.
In the illustrative example, the data point with the larges value for $X$ (33 in \Cref{fig:sampling}) defines the upper end of transportability. 

\section{Conclusion}
\label{sec:conclusion}

Transportability methods have the potential to improve the external validity of results from controlled experiments and increase their practical relevance.
If covariates moderate the ATE of a phenomenon of interest while simultaneously affecting trial eligibility, external validity is under threat.
However, if observational data about those covariates from a larger sample is available, results can be transported to this larger sample using transportability methods.
Their application could help address several long-standing issues with experimentation in SE.
Still, the path to adopt transportability methods in SE requires addressing several challenges in order to meet all preconditions.
Targeting this goal will encourage SE researchers to explore and understand relevant covariates, collect data about them, and actively reason about the representativeness of their experimental subjects, objects, and tasks.
In future work, we aim to demonstrate the application to real cases of SE research.


\bibliographystyle{ACM-Reference-Format}
\bibliography{material/references} 

\end{document}